\documentclass[aps,prb,floats,twocolumn,10pt]{revtex4-1}
\usepackage{graphicx}
\usepackage{amssymb}
\usepackage{amsmath}

\usepackage{color}
\usepackage[dvipsnames]{xcolor}
\usepackage{slantsc}
\usepackage{float}
\graphicspath{{./FIG_FINALI/}}
\usepackage{bbold}
\usepackage[caption=false]{subfig}
\usepackage{array}
\usepackage{braket}
\usepackage{scrextend}
\usepackage{hyperref}
\hypersetup{backref,pdfpagemode=FullScreen,colorlinks=true,allcolors=blue}
\usepackage{leftidx}

\begin{document}
\title{XDFT: an efficient first-principles method for neutral excitations in molecules}
\author{Subhayan Roychoudhury}
\affiliation{School of Physics, AMBER and CRANN Institute, Trinity College Dublin, Dublin 2, Ireland}
\author{Stefano Sanvito}
\affiliation{School of Physics, AMBER and CRANN Institute, Trinity College Dublin, Dublin 2, Ireland}
\author{David D. O'Regan}
\email{david.o.regan@tcd.ie}
\affiliation{School of Physics, AMBER and CRANN Institute, Trinity College Dublin, Dublin 2, Ireland}

\begin{abstract}
State-of-the-art methods for calculating neutral excitation energies are typically demanding and limited to 
single electron-hole pairs and their composite plasmons. Here we introduce {\it excitonic density-functional
theory} (XDFT) a computationally light, generally applicable, first-principles technique for calculating neutral excitations
based on generalized 
constrained DFT. 
In order to simulate an $M$-particle excited state
of an $N$-electron system, 
XDFT automatically optimizes a constraining potential to 
confine $N-M$ electrons within the ground-state Kohn-Sham valence subspace. We demonstrate the efficacy of XDFT by calculating the lowest 
single-particle singlet and triplet excitation energies of the well-known Thiel molecular test set, with results which are  
in excellent agreement with time-dependent DFT. Furthermore, going beyond the capability of adiabatic 
time-dependent DFT, we show that XDFT can successfully capture  double excitations. Overall our
method makes optical gaps, excition bindings and oscillator strengths readily accessible at a computational
cost comparable to that of standard DFT. As such, XDFT appears as an ideal candidate to work 
within high-throughput discovery frameworks and 
within linear-scaling 
methods for large systems.
\end{abstract}

\maketitle
The first principles calculation of excited-state energies of quantum systems holds crucial importance 
for the study of solar cells~\cite{doi:10.1021/ar500089n}, organic light emitting diodes~\cite{1708.05247}, and
chromophores in biological systems~\cite{PhysRevLett.90.258101},  to name but a few. With some exceptions
density-functional theory (DFT), which is the primary {\it ab initio} workhorse for computing ground state 
properties~\cite{Hasnip20130270,hafner_wolverton_ceder_2006}, typically falls short on such tasks, although 
efforts are underway to extend the foundation of DFT to excited states~\cite{PhysRevB.31.6264,PhysRevLett.83.4361,PhysRevA.80.012508,PhysRevA.59.3359,PhysRevA.54.3912,1742-6596-388-1-012011}. The most commonly used first-principles method for calculating excitation 
energies, at least of finite systems, is perturbative time-dependent density functional theory (TDDFT)~\cite{PhysRevLett.52.997,doi:10.1142/9789812830586_0005,Fundamentals_of_TDDFT}. However, TDDFT 
has two significant limitations: 1) its considerable computational costs, which severely limit the size of the
systems that it can investigate~\cite{doi:10.1063/1.4926837}, and 2) its inability to treat double (two-electron) or higher-order
excitations 
 within 
adiabatic approximations to the exchange-correlation (XC)
kernel~\cite{doi:10.1063/1.1651060,ELLIOTT2011110,doi:10.1063/1.4953039}. Over the years several 
first-principles schemes based on time-independent DFT have been developed for 
calculating neutral excitation energies, such as ensemble 
DFT~\cite{QUA:QUA560382470,PhysRevLett.119.033003,PhysRevB.95.035120}, restricted open-shell 
Kohn-Sham DFT~\cite{FILATOV1999429,doi:10.1063/1.475804,OKAZAKI1998109,doi:10.1063/1.4801790}, 
constricted variational DFT~\cite{doi:10.1063/1.3114988,doi:10.1021/ct300372x}, 
$\Delta$SCF-DFT~\cite{doi:10.1063/1.2977989,doi:10.1063/1.3530801} and the 
maximum overlap method~\cite{doi:10.1021/jp801738f,doi:10.1063/1.4789813}. 
All of these approaches have their strengths and weaknesses in terms 
of computational costs and ease of both implementation and convergence. XDFT, like some of the latter methods, is motivated by the existence of a variational DFT, with a minimum 
principle and an equivalent non-interacting Kohn-Sham (KS) state, for an individual excited state of  interacting electrons~\cite{PhysRevB.31.6264,PhysRevLett.83.4361,doi:10.1063/1.2977989}. 
We refer the reader to 
Ref.~(\onlinecite{doi:10.1146/annurev-physchem-032511-143803}) for a recent review
of TDDFT, and to 
Ref.~(\onlinecite{doi:10.1063/1.3114988})
for a foundational comparison between  DFT-based variational approaches and  TDDFT. 

A neutral excitation, within the quasiparticle picture, is the promotion of one or more electrons from occupied 
levels to empty ones, resulting in the creation of bound electron-hole pairs, or excitons, and  consequent 
energy-level relaxation. In this Letter, we introduce \textit{excitonic DFT} (XDFT), an inexpensive, fully 
first-principles method based on constrained DFT 
(cDFT)~\cite{PhysRevLett.53.2512,KadukB_KT_VT,WVV} for calculating neutral excitation energies in finite 
systems such as molecules and clusters. XDFT scales with the atom count $N$ as per ground-state DFT, namely as $\mathcal{O}(N^3)$. This contrasts with 
methods like TDDFT, which typically scales as $\mathcal{O}(N^4)$~\cite{JCC:JCC8} and the Bethe-Salpeter 
equation (BSE), which goes as $\mathcal{O}(N^6)$~\cite{PhysRevB.92.075422}. In addition, unlike 
TDDFT and BSE, which are highly memory intensive, XDFT has a memory overhead comparable to that of 
standard DFT. 
Crucially, it  avoids the calculation
of unoccupied KS orbitals entirely.
Therefore, in terms of computational efficiency, XDFT offers significant advantages over other
methods and appears to be readily compatible with
 high-throughput frameworks, the study of large-systems, 
and KS methods beyond DFT.

Much like cDFT~\cite{KadukB_KT_VT,WVV}, XDFT searches for the ground-state energy of a system subject to confining a given 
number of electrons with spin $\sigma$, $N^{\sigma}_c$, to a desired subspace. Such a constraining condition 
may be written as 
\begin{align}
{\rm Tr}\big[\hat{\rho}^{\sigma} \hat{\mathbb{P}} \big]=N^{\sigma}_c\;,
\end{align}
where `Tr' denotes the trace, $\hat{\rho}^{\sigma}$ is the spin-dependent Fermionic density operator and 
$\hat{\mathbb{P}}$ is a projection operator onto the desired subspace. Then, the ground state of the system 
subject to the constraint is found at the stationary point of the functional
\begin{align}
W[\hat{\rho},V_c]=E[\hat{\rho}]+V_c\left({\rm Tr}\big[\hat{\rho}^{\sigma}\hat{\mathbb{P}}\big]-N^{\sigma}_c\right)\:,
\end{align}
where $V_{\textit{c}}$ is a Lagrange multiplier. For a fixed $V_c$, the second term on the right-hand side serves 
to modify the ground state potential by adding the term $V_c\hat{\mathbb{P}}$. One then minimizes $W[\hat{\rho},V_\textit{c}]$ 
with respect to $\hat{\rho}$, just as $E[\hat{\rho}]$ in regular DFT. At the $V_\textit{c}$-dependent minima $W[\hat{\rho},V_\textit{c}]$ can be thought of 
as a function~\cite{WVV}, $W(V_\textit{c})$, of $V_\textit{c}$. 
The maxima of $W(V_\textit{c})$ occur at stable states of 
 the constrained system~\cite{ORT}, at which the
 value of $W$ is the  total energy of interest.

In conventional cDFT, the subspace spanned by $\hat{\mathbb{P}}$ is a spatial region. If $\hat{\mathbb{P}}$  spans two spatial 
regions with opposite weighting, then one can enforce a
charge-separated density configuration for the simulation of charge-transfer excitations~\cite{doi:10.1021/ct0503163,PhysRevB.88.165112,C6CP00528D, doi:10.1021/acs.jctc.6b00815, doi:10.1021/acs.jctc.6b00564}. 
In this work, in order to access excitations beyond charge-separated states, we lift the restriction of the spatial 
confinement by defining the cDFT subspace in terms of the KS eigenstates. 
For a neutral $N$-electron system XDFT locates the energy of the lowest $M$-electron excited state by confining 
$ N-M $ electrons within the  valence KS subspace of the unconstrained DFT ground-state. 
This circumvents 
the need for any prior, empirical specification of 
subspaces as in conventional cDFT.
The projector that defines XDFT is 
\begin{align*}
\hat{\mathbb{P}}=\hat{\rho}_0=\sum_i f_i\ket{\psi_i}\bra{\psi_i}\:,
\end{align*}
where $\hat{\rho}_0$ is the ground state density operator, $\ket{\psi_i}$ is the $i^\textrm{th}$ KS orbital and $f_i$ is its 
occupation number. 
Once the energy of the first 
excited state, $ W$, is determined by optimizing
$V_\textit{c}$, then the lowest excitation energy, $E^*$, can be evaluated as a total 
energy difference from the ground state DFT energy, $E_0$, namely as $E^*=W-E_0$.

XDFT is formally an orbital-dependent DFT, and its energy is separately invariant under arbitrary unitary 
transformations among the occupied Kohn-Sham orbitals 
of  the ground-state and of 
the constrained ground-state.
The KS wave-function of the  excited state obtained 
from XDFT is orthonormal to that of 
the ground state. This is because it is a Slater determinant (SD) composed of  KS 
orbitals the  highest  of which is, as a result of the 
constraint definition, orthonormal to each of the KS  
orbitals comprising the  ground state KS  wave-function.

\begin{figure}
\centering
\includegraphics[width=0.7\columnwidth]{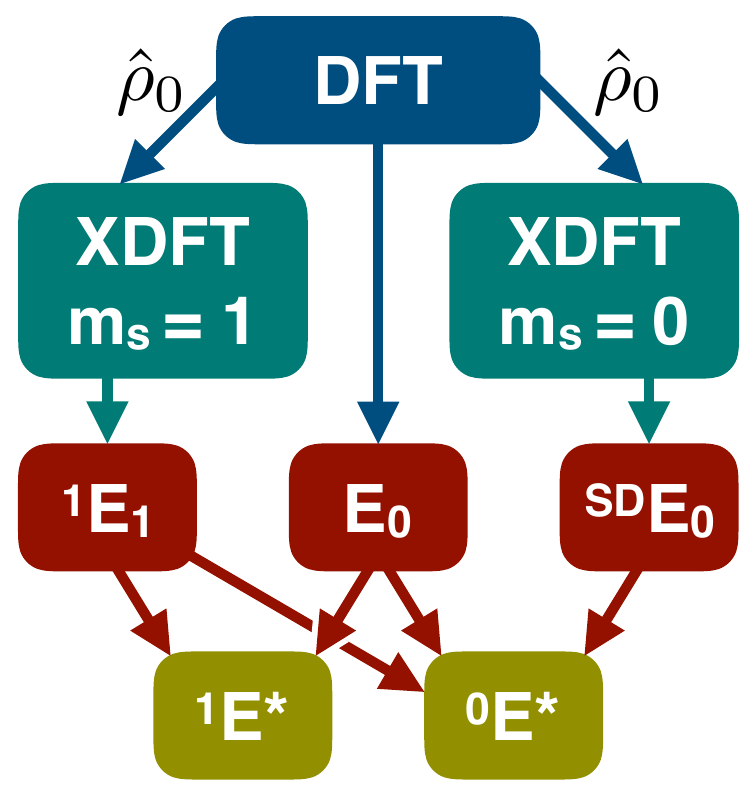}
\captionsetup{justification=justified,singlelinecheck=false}
\caption{XDFT flowchart for calculating singlet, $^{S=0}E^\ast$, and triplet, $^{S=1}E^\ast$, 
excitation energies. The ground state density operator obtained from a DFT calculation, $\hat{\rho}_0$, is used 
to define the cDFT constraint. The energies obtained from DFT, $E_0$, and from XDFT, $^{S=1}E_{m_s=1}$
and $^{\textrm{SD}}E_{m_s=0}$, are then used to find the excitation energies using 
Eqs.~(\ref{SingletEnergySumMethod}),  (\ref{EnergyDifferenceForOpticalGap1}), 
and (\ref{EnergyDifferenceForOpticalGap2}).}
\label{FIG:Schematic}
\end{figure}

XDFT can be used to simulate combinations of charge and spin excitations. Given a closed-shell ground state, triplet single-electron excitations and singlet double excitations are
 straightforward to access with 
a single constraint. These both
incur the cost of just two DFT calculations 
-- the ground-state one and the constrained one.
In both cases, the electron-promotion
constraint can be applied to 
the sum of the density operators for each spin, 
and the triplet state can be selected by setting  
$m_s = 1$. Slightly more work is required to access singlet single
and triplet double excitations, as we now discuss. 

Given a closed-shell ground state, 
the final state of a singlet single 
excitation can not be represented by a single SD, 
and so the corresponding excitation energy 
cannot be obtained straightforwardly 
from a single cDFT calculation.
Fortunately, for the non-interacting KS system both the closed-shell singlet excited state $^{S=0}\Psi^{\textrm{KS}}_{m_s=0}$ (with a non-interacting energy $^{S=0}E^{\textrm{KS}}_{m_s=0}$) and the open-shell singlet 
excited state $^{S=1}\Psi^{\textrm{KS}}_{m_s=0}$ (with $^{S=1}E^{\textrm{KS}}_{m_s=0}$) 
can be written out as a linear combination of 
the same pair of SDs, within a frozen-orbital treatment. 
These two SDs are then degenerate, with a non-interacting
energy $\leftidx{^{\textrm{SD}}}{E}{^{\textrm{KS}}_{m_s=0}}$. 
Invoking the multiplet sum  
method~\cite{CRAMER1995165,doi:10.1146/annurev-physchem-032511-143803}, 
we  can thereby express the non-interacting energy of a  
closed-shell singlet state approximately as
\begin{align}\label{LinearCombEn}
^{S=0}E^{\textrm{KS}}_{m_s=0}=2\times                         \leftidx{^{\textrm{SD}}}{E}{^{\textrm{KS}}_{m_s=0}} - ^{S=1}E^{\textrm{KS}}_{m_s=0}\:.
\end{align}
In order to access one of these degenerate SDs that
make up the singlet in practice, we apply the XDFT 
constraint to one spin channel only, using $M=1$ and $m_s=0$. 
At this point, we  make a final assumption that
Eq.~(\ref{LinearCombEn}) may be used to approximately
evaluate the energy of the interacting system.
Keeping in mind that the three triplet states for 
$m_s=-1,0,1$ are degenerate, we arrive at 
\begin{align}\label{SingletEnergySumMethod}
^{S=0}E_{m_s=0}\approx 2\times \leftidx{^{\textrm{SD}}}{E}{_{m_s=0}} - ^{S=1}E_{m_s=1}.
\end{align}
The advantage of Eq.~(\ref{SingletEnergySumMethod}) is that it  involves only the
energies of two single-SD states that  are available
using XDFT. This approximation has the same regime
of validity as the frozen-orbital approximation
for the KS orbitals.
Each  term  in Eq.~(\ref{SingletEnergySumMethod}) 
derives from an interacting system that is
obtained from an equivalent unrestricted KS system having the same density and spin density. Ultimately, the task of 
determining the triplet and singlet single excitation energies reduces to running the following three first-principles 
calculations (see Fig.~\ref{FIG:Schematic}):
\begin{enumerate}
\item A DFT calculation to determine the ground-state energy $E_0$ and density operator $\hat{\rho}_0$.
\item An XDFT calculation with $m_s=1$, confining $N-1$ electrons to the total valence subspace of the 
DFT run. This gives the energy 
of the lowest-lying interacting triplet  state, $^{S=1}E_{m_s=1}$.
\item An XDFT calculation with $m_s=0$,  confining $(N/2)-1$ electrons to the spin-up valence subspace of the DFT 
run, to find the energy $^{\textrm{SD}}E_{m_s=0}$.
\end{enumerate}
Next, we use Eq.~(\ref{SingletEnergySumMethod}) to approximate the energy $^{S=0}E_{m_s=0}$ of the 
 excited singlet. Finally, we calculate
the triplet and  singlet 
neutral gaps,  respectively, as
\begin{align}\label{EnergyDifferenceForOpticalGap1}
^{S=1}E^\ast&{}=^{S=1}E_{m_s=1}-E_0 \quad \mbox{and} \\
^{S=0}E^\ast&{}=^{S=0}E_{m_s=0}-E_0 \; .
 \label{EnergyDifferenceForOpticalGap2}
\end{align} 

We have implemented the XDFT formalism in the linear-scaling first-principles code {\sc onetep}~\cite{CKSPDHAAMMCP}, 
which  variationally optimizes a minimal set of localized, non-orthogonal generalized Wannier Functions (NGWF), expanded in terms of 
psinc functions~\cite{doi:10.1063/1.1613633,PhysRevB.66.035119}, to minimize the total energy. 
{\sc onetep} is equipped with an automated conjugate-gradients method for updating the cDFT (or XDFT) Lagrange 
multiplier~\cite{PhysRevB.93.165102,ORT,1802.01669}. 
We have used this, together with the Perdew-Burke-Ernzerhof 
(PBE) XC functional~\cite{PhysRevLett.77.3865} to calculate the lowest singlet excitation energies of 
the 28 closed-shell organic molecules contained in the well-known Thiel set~\cite{doi:10.1063/1.2973541}. Our 
calculations are performed using scalar relativistic norm-conserving pseudopotentials, a plane-wave cutoff energy of $1500$~eV 
and a radius of $14.0$~a$_0$ for the NGWFs. 
The Martyna-Tuckerman periodic boundary correction
scheme~\cite{MT} was used with a parameter of 7.0~$a_0$.
The constrained KS system  was found to contain
symmetry-protected partial-filling of degenerate 
highest occupied state
in certain molecules, and so we used  
finite-temperature ensemble DFT as implemented in {\sc onetep}~\cite{ARSCKS} in all cases.

\begin{figure}
\centering
\includegraphics[width=0.9\columnwidth]{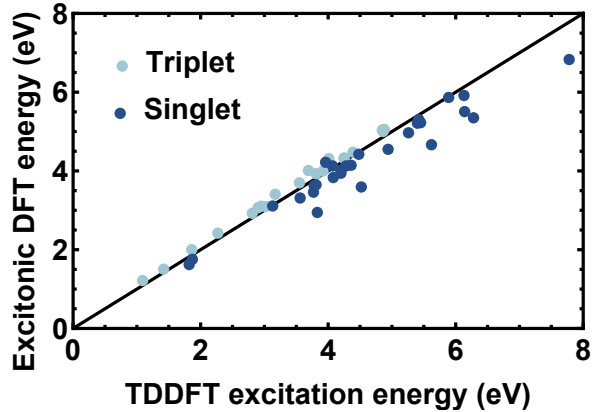}
\captionsetup{justification=justified, singlelinecheck=false}
\caption{(Color online) The lowest excitation energies of molecules belonging to the Thiel 
set~\cite{doi:10.1063/1.2973541} obtained with XDFT and with adiabatic linear-response TDDFT (from 
Refs.~[\onlinecite{doi:10.1021/ct900298e}] and~[\onlinecite{doi:10.1021/ct100005d}]). The 
PBE~\cite{PhysRevLett.77.3865} XC functional has been used in both cases. The dark and 
the light dots denote singlet and triplet gaps, respectively. The diagonal line indicates perfect
agreement between XDFT and TDDFT.}
\label{FIG.OG_Sing_Trip}
\end{figure}

In Fig.~\ref{FIG.OG_Sing_Trip}, we show a scatter plot of the singlet and triplet excitation energies calculated 
with XDFT against those obtained with linear-response
TDDFT and 
adiabatic PBE in Ref.~[\onlinecite{doi:10.1021/ct900298e}] 
(singlets) and Ref.~[\onlinecite{doi:10.1021/ct100005d}] (triplets). The TDDFT results are generally in agreement 
with experimental values (see the supporting information in Ref.~[\onlinecite{doi:10.1063/1.2889385}]). The triplet 
energies show an excellent agreement with TDDFT because, being SDs, the KS triplet excited 
states are directly accessible using a single XDFT calculation. The figure also demonstrates that,
in spite of the multiplet sum approximation, 
XDFT calculates  yields singlet energies with 
a remarkably good accuracy.

\begin{figure}
\centering
\includegraphics[width=\columnwidth]{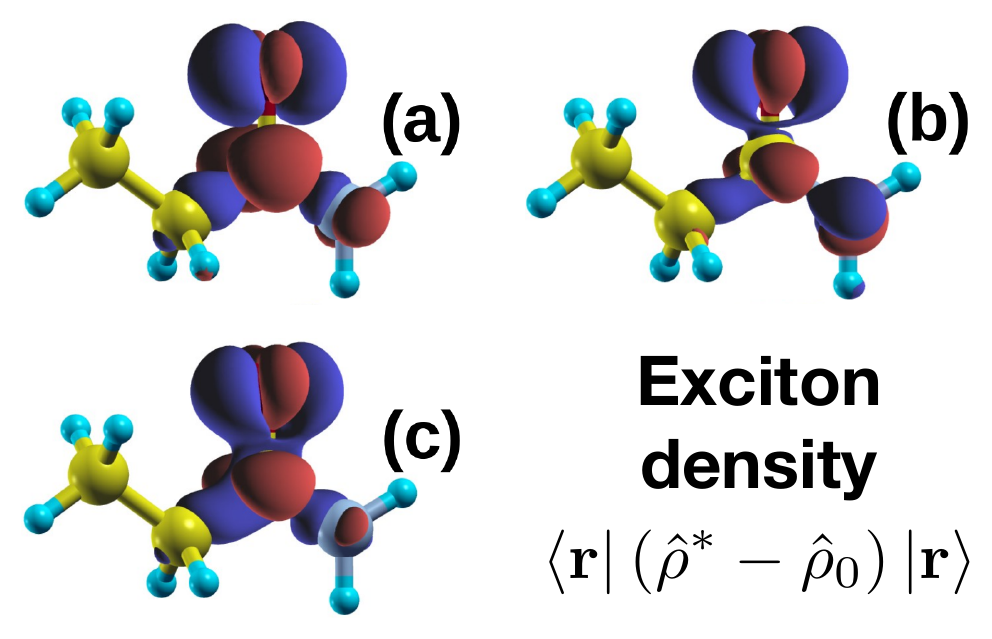}
\caption{Exciton charge density for the propanamide molecule, calculated with an isosurface value 
of $\pm 0.05$~e\AA$^{-3}$. Panel (a) shows the charge-density
difference between the ground-state KS 
{\sc lumo} and {\sc homo} orbitals, while (b) and (c) show, respectively, the singlet and triplet 
exciton densities generated as a difference between the XDFT
and DFT total densities.}
\label{Fig:ExcitonDensityPlot}
\end{figure}

A plot of the difference in charge density between the excited state and the ground state provides a good 
approximation to the exciton charge density. In Fig.~\ref{Fig:ExcitonDensityPlot} we show such plots for a 
representative molecule of the Thiel set, propanamide. 
Fig.~\ref{Fig:ExcitonDensityPlot}(a) shows an 
 approximation to the exciton density based on the ground-state
 KS orbitals, which neglects the orbital
 relaxation and exciton binding.
Since it captures these effects, the singlet (b) and triplet (c)
isosurfaces generated using XDFT
(and, in the case
of the singlet, the multiplet sum method applied to the
total electron densities)
reflect a greater degree of exciton density
localization than (a). 
Due to Pauli exclusion, furthermore, the singlet (b)
exciton density attains a greater spatial localization
than the triplet one (c).

In tests on excitations for which  
non-adiabatic linear-response TDDFT is known
to perform well,  XDFT with $m_s=1$ yields 
the precisely the same total energy as an 
unconstrained DFT calculation with $m_s=1$
(i.e., DFT with a fixed spin-moment of $1~\mu_\textrm{B}$).
Similarly, it does not affect the total energy if we apply
an additional
constraint to the spin population within the
ground-state KS manifold.
Based on this evidence, it appears that fixing
the cDFT subspace to the ground-state 
KS valence manifold does not 
introduce any appreciable approximation to the 
definition of an excited state, at least in this 
`adiabatic' regime. This is notwithstanding
the fact that KS wave-function orthonormality
does not imply the orthonormality of interacting 
wave-functions.

Before concluding our discussion on single-electron 
excitations, we note that higher-energy 
excitations can be simulated in XDFT by employing multiple constraints. For example, if the valence subspaces 
of the ground state and the first XDFT excited state are projected onto by $\hat{\mathbb{P}}_0$ and $\hat{\mathbb{P}}_1$, 
respectively, then the energy of the second  excited state can be found by confining $N-1$ electrons 
within the subspace of $\hat{\mathbb{P}}_0$ using a Lagrange multiplier $V^1_\textit{c}$ and, separately, confining $N-1$ 
electrons within the subspace of $\hat{\mathbb{P}}_1$ using a  multiplier $V^2_\textit{c}$. In general, 
the total-energy of the $I^\textrm{th}$ excited state system
of a given spin symmetry will be found at the stationary point of 
\begin{align}
W = E \left[ \hat{\rho} \right] + \sum_i^I V_\textit{c}^i \left( \mathrm{Tr} \left[ \hat{\rho} \hat{\mathbb{P}}_{i-1} \right] - \left( N - 1\right)  \right)\:.
\end{align}

Finally, we explore the ability of XDFT to calculate energies of     excitations with strong
double (two-electron) character, which are considered to be inaccessible to adiabatic TDDFT~\cite{doi:10.1063/1.1651060,ELLIOTT2011110,doi:10.1063/1.4953039}.
This is the case by construction within the linear response regime, but not necessarily so within full-response TDDFT.
The XDFT method is non-perturbative, in that the Hartree
and XC potentials are calculated
self-consistently with the density in the excited state.
Thus, XDFT is not limited to single excitations.
In the  benchmark case of atomic 
beryllium, the first double excitation promotes two electrons from the $2s$ to the $2p$ 
orbitals~\cite{A910321J}. We calculated the singlet and triplet double excitations of Be by means of (the 
$1s$ electrons were pseudized, rendering multiplet summation
unnecessary):
\begin{enumerate}
\item A ground-state DFT calculation.
\item An XDFT calculation with $m_s = 0$, confining $0$ electrons to the total valence  subspace of the DFT run. This gives the
energy $^{S=0}E_{m_s=0}^2$ of the lowest-lying interacting doubly-excited singlet state.
\item An XDFT calculation with $m_s=1$,  confining $0$ electrons to the total valence  subspace of the DFT run.  This yields the energy $^{S=1}E_{m_s=1}^2$.
\end{enumerate}

\begin{figure}[b]
\centering
\includegraphics[width=0.8\columnwidth]{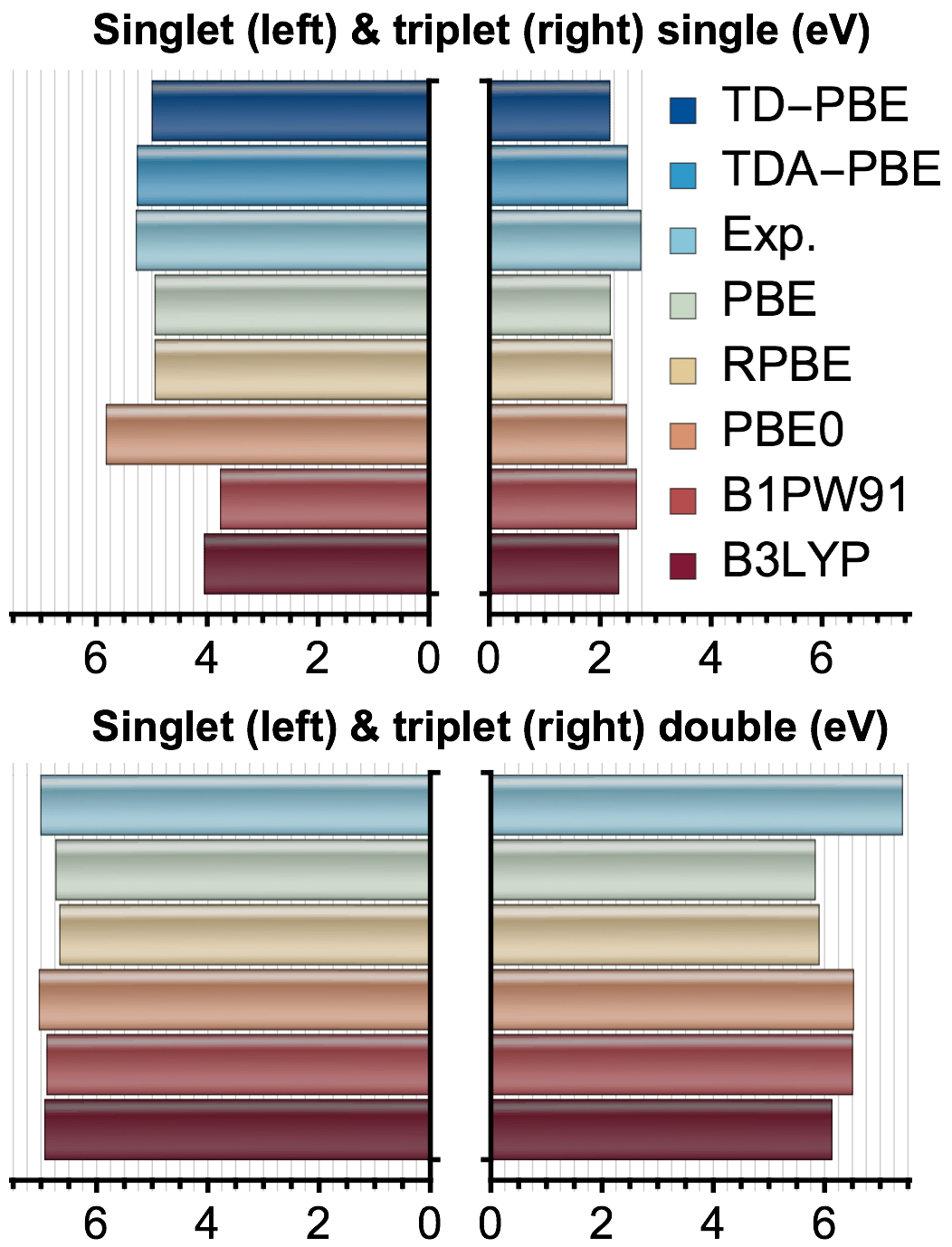}
\captionsetup{justification=justified,singlelinecheck=false}
\caption{The excitation energies of atomic Be from experiment, linear-response TDDFT (the \textsc{ONETEP} linear-scaling implementation~\cite{doi:10.1063/1.4817330,doi:10.1063/1.4936280}), and XDFT. The top panels show  single-electron 
excitation energies and the bottom panels show double (two-electron) excitation energies. 
The functionals tested were the generalized gradient approximation parameterizations PBE~\cite{PhysRevLett.77.3865} and 
RPBE~\cite{PhysRevB.59.7413}, and the hybrid functionals   PBE0~\cite{Adamo19996158}, 
B3LYP~\cite{doi:10.1063/1.464304}, and B1PW91~\cite{ADAMO1997242}.
For the single excitations, we have included the results of 
linear-response TDDFT calculations within adiabatic PBE, 
where double excitations are inaccessible. TDA here
refers to the same calculations but within the Tamm-Dancoff approximation~\cite{HIRATA1999291}.
We also provide experimental 
values taken from Ref.~[\onlinecite{doi:10.1063/1.555999}].}
\label{FIG:BeExcitationEnergyPlot}
\end{figure}

The singlet and triplet double excitation energies were   obtained  as 
$\leftidx{^{S=0}}{E}{^{2 \ast}} = \leftidx{^{S=0}}{E}{^2_{m_s=0}}-E_0$ and 
$\leftidx{^{S=1}}{E}{^{2 \ast}} = \leftidx{^{S=1}}{E}{^2_{m_s=1}}-E_0$, respectively.  In Fig.~\ref{FIG:BeExcitationEnergyPlot} we plot the single and 
double excitation energies of Be calculated with semi-local
and hybrid XC-functionals. These  
agree well with those calculated with ensemble DFT in Ref.~[\onlinecite{PhysRevLett.119.033003}], for all four 
excitation types. 
The singlet single-electron PBE excitation energy is also in very close
agreement with our own linear-response TDDFT(PBE) result, 
indicating that the multiplet sum approximation is successful in this system.
We note, however, that while our singlet 
double excitation energies agree well with experimental values, 
this is much less the case for our triplet double energies. Experimentally, the singlet $2s^2\rightarrow 2p^2$ 
excitation is lower in energy than the triplet one, and this has
been explained as resulting from a mixing of the singlet
double with higher singlet single excitations~\cite{doi:10.1063/1.555999}.
Our results would support the opposite conclusion, 
however, 
since it is the  triplet state which is poorly described.
XDFT is capable of accessing excitations
of non-integer electron character (e.g, mixed single and double
excitations) with the aid of ensemble DFT~\cite{ARSCKS,MVP}, 
in principle, and
this is a promising avenue for future investigation. 

In summary, we introduce the XDFT method for calculating the excited-state energies 
of finite systems by means of a small number of coupled
DFT calculations.
XDFT generalizes constrained DFT, in essence, 
by removing the necessity for users to pre-define the targeted
subspaces.
Unlike linear-response TDDFT or BSE, no reference is made
to unoccupied orbitals.
XDFT closely reproduces the TDDFT values for triplet and also, 
with the help of an additional approximation, in most
cases the singlet excitation energies 
of the Thiel molecular test set.
Unlike adiabatic TDDFT, however, XDFT can readily 
access the energies of double excitations, effectively
circumventing the requirement for non-adiabaticity in TDDFT.
We demonstrate this in a successful application to the 
well-known test case of the beryllium atom.

This work is supported by the European Research Council project {\sc quest}. 
We acknowledge G. Teobaldi and N. D. M. Hine for their implementation  of cDFT in \textsc{ONETEP}.
We acknowledge the 
DJEI/DES/SFI/HEA Irish Centre for High-End Computing (ICHEC) and Trinity Centre for High Performance 
Computing (TCHPC) for  the provision of computational resources. 


%

\end{document}